\documentclass[10pt,letterpaper,twocolumn]{article} %% two column, final layout
\usepackage{ol2} %% do not use with REVTeX4
\usepackage{hyperref,cite}

\usepackage{graphicx,ulem}
\usepackage{amsmath,amstext}

\newcommand{\Ein}{{\mathcal E}_0}%{1,\rm in}}
\newcommand{\neffs}{N_{\rm eff}}%
\newcommand{\nshgb}{N_{\rm SHG}}%
\newcommand{\nkerrb}{N_{\rm Kerr}}%
\newcommand{\ld}{L_{\rm D,1}}
\newcommand{\deff}{d_{\rm eff}}
\newcommand{\sgn}{{\rm sgn}}

\newcommand{\lcoh}{L_{\rm coh}}
\newcommand{\kp}{k^{(1)}}
\newcommand{\kpp}{k^{(2)}}
\newcommand{\tin}{T_0}%{T_{1,\rm in}}

\begin{document}
\twocolumn[ 
\title{Nonlocal explanation of stationary and nonstationary regimes in
  cascaded soliton pulse compression} 
% \author{M. Bache} 
% \affiliation{COM$\bullet$DTU,
%   Technical University of Denmark, DK-2800 Lyngby, Denmark}
% \email[Corresponding author: ]{bache@com.dtu.dk}
% \author{O. Bang} 
% \affiliation{COM$\bullet$DTU,
%   Technical University of Denmark, DK-2800 Lyngby, Denmark}
% \author{J. Moses} 
% \affiliation{Optics and Quantum Electronics Group, Massachusetts Institute of
%  Technology, Cambridge, MA 02139}
% \author{F. W. Wise} 
% \affiliation{Department of Applied and Engineering Physics, Cornell
%   University, Ithaca, New York 14853}
\author{M. Bache,$^{1,*}$ O. Bang,$^1$ J. Moses,$^2$ and F. W. Wise$^3$}
\affiliation{$^1$COM$\bullet$DTU, 
%Department of Communications, Optics \&  Materials,
Technical University of Denmark,
  Building. 345v, DK-2800 Lyngby, Denmark\\
$^2$Research Laboratory of Electronics, Massachusetts Institute of
Technology, Cambridge, Massachusetts 02139, USA\\ 
$^3$Department of Applied and Engineering Physics, Cornell
  University, Ithaca, New York 14853, USA\\
$^*$Corresponding author: bache@com.dtu.dk}
\begin{abstract}
% We study pulse compression in quadratic materials and show that
% group-velocity mismatch creates two nonlocal oscillatory and localized
% regimes that describe nonstationary and stationary pulse
% compression. The theory is used to accurately predict the transition. 
  We study soliton pulse compression in materials with cascaded
  quadratic nonlinearities and show that the group-velocity mismatch
  creates two different temporally nonlocal regimes. They correspond
  to what is known as the stationary and nonstationary regimes. The
  theory accurately predicts the transition to the stationary regime,
  where highly efficient pulse compression is possible.
\end{abstract}
\maketitle
]

Efficient soliton pulse compression is possible using second-harmonic
generation (SHG) in the limit of large phase mismatch, because a
Kerr-like nonlinear phase shift is induced on the fundamental wave
(FW). Large negative phase shifts can be created, since the phase
mismatch determines the sign and magnitude of the effective cubic
nonlinearity.  This induced \textit{self-defocusing} nonlinearity thus
creates a negative linear chirp through an effective self-phase
modulation (SPM) term, and the pulse can therefore be compressed with
normal dispersion. Beam filamentation and other problems normally
encountered due to \textit{self-focusing} in cubic media are therefore
avoided.  This self-defocusing soliton compressor can create
high-energy few-cycle fs pulses in bulk materials with no power limit
\cite{liu:1999,ashihara:2002,bache:2007,moses:2006}.  However, the
group-velocity mismatch (GVM) between the FW and second harmonic (SH)
limits the pulse quality and compression ratio.  Especially very short
input pulses ($< 100$ fs) give asymmetric compressed pulses and pulse
splitting occurs \cite{ilday:2004,moses:2006}. In this case, the
system is in the \textit{nonstationary regime}, and conversely when
GVM effects can be neglected it is in the \textit{stationary regime}
\cite{ilday:2004,moses:2006,bache:2007}.  Until now, the stationary
regime was argued to be when the characteristic GVM length is 4 times
longer than the SHG coherence length \cite{liu:1999}, while a more
accurate perturbative description showed that the FW has a GVM-induced
Raman-like term \cite{ilday:2004,moses:2006}, which must be small for
the system to be in the stationary regime \cite{moses:2006}.  However,
no precise definition of the transition between the regimes exists.
%the stationary and nonstationary regimes exists.

On the other hand, the concept of nonlocality provides accurate
predictions of quadratic spatial solitons
\cite{nikolov:2003,larsen:2006}, and many other physical systems (see
\cite{krolikowski:2004} for a review).  Here we introduce the concept
of nonlocality to the temporal regime and soliton pulse compression in
quadratic nonlinear materials. As we shall show, GVM, the phase
mismatch, and the SH group-velocity dispersion (GVD) all play a key
role in defining the nonlocal behavior of the system. Two different
nonlocal response functions appear naturally, one with a localized
amplitude -- representing the stationary regime -- and one with a
purely oscillatory amplitude -- representing the nonstationary regime.
In the presence of GVM they are asymmetric and thus give rise to a
Raman effect on the compressed pulse. 

In the theoretical analysis we may neglect diffraction,
higher-order dispersion, cubic Raman terms, and self-steepening to get
the SHG propagation equations for the FW ($\omega_1$) and SH
($\omega_2=2\omega_1$) fields $E_{1,2}(z,t)$
\cite{moses:2006b,bache:2007}:  
%describing SHG in
%quadratic nonlinear materials are
%\begin{subequations}
\begin{align}
  \label{eq:shg-bulk-fh-no-kerr-dim}
(i\partial_z-\tfrac{1}{2}\kpp_1&\partial_{tt})E_1+\kappa_1
    E_1^*E_2e^{i\Delta k z}
\\\nonumber  &
+\rho_1E_1\left(|E_1|^2+\eta|E_2|^2\right)=0,
 \\\nonumber  
 (i\partial_z-id_{12}\partial_{t}-\tfrac{1}{2}\kpp_2&\partial_{tt})
     E_2+\kappa_2E_1^2e^{-i\Delta k z} 
 \\&
 +\rho_2E_2\left(|E_2|^2+\eta|E_1|^2\right)=0,
   \label{eq:shg-bulk-sh-no-kerr-dim}
\end{align}
where $\eta=2/3$ for Type I SHG \cite{bache:2007}. $\kappa_j=\omega_1
\deff/cn_j$, $\deff$ is the effective quadratic nonlinearity,
$\rho_j=\omega_jn_{{\rm Kerr},j}/c$, and $n_{{\rm Kerr},j}=3{\rm
  Re}(\chi^{(3)})/8 n_j$ is the cubic (Kerr) nonlinear refractive
index. The phase mismatch is $\Delta k=k_2-2k_1$, with
$k_j=n_j\omega_j/c$. $n_j$ is the refractive index, and
$k_j^{(n)}=\partial^n k_j/\partial \omega^n|_{\omega=\omega_j}$
accounts for dispersion. The time coordinate moves with the FW group
velocity $v_{\rm g,1}=1/\kp_1$, giving the GVM term $d_{12}=v_{\rm
  g,1}^{-1}-v_{\rm g,2}^{-1}$. In dimensionless form, $\tau=t/\tin$,
where $\tin$ is the FW input pulse duration, $\xi=z/\ld$, where
$\ld=\tin^2/|\kpp_1|$ is the FW dispersion length, and $U_1=E_1/\Ein$,
$U_2=E_2/\sqrt{\bar{n}}\Ein$, where $\Ein=E_1(0,0)$ and $\bar
n=n_1/n_2$.  Equations~(\ref{eq:shg-bulk-fh-no-kerr-dim}) and
(\ref{eq:shg-bulk-sh-no-kerr-dim}) become \cite{bache:2007}
%\begin{subequations}\label{eq:shg-dimless}
% \begin{align}
%   \label{eq:shg-bulk-fh-no-kerr}
%   i\frac{\partial \phi_1}{\partial \xi}-D_1
%   \frac{\partial^2 \phi_1}{\partial \tau'^2}
%   +\phi_1^*\phi_2e^{i\Delta\beta \xi}&=0\\
%   \label{eq:shg-bulk-sh-no-kerr}
%   i\frac{\partial \phi_2}{\partial \xi}-i\delta
%   \frac{\partial \phi_2}{\partial \tau'}
%   -D_2 \frac{\partial^2 \phi_2}{\partial \tau'^2}
% %&\\
%   +\phi_1^2e^{-i\Delta\beta \xi}&=0
% \end{align}
\begin{align}
  \label{eq:shg-bulk-fh-no-kerr}
  &(i\partial_\xi 
  -D_1\partial_{\tau\tau}) U_1
  +\sqrt{\Delta \beta}\nshgb U_1^*U_2e^{i\Delta\beta \xi}
\nonumber\\
&
\phantom{i\partial_\xi-D_1}
+\nkerrb^2U_1\left(|U_1|^2+\eta\bar{n}|U_2|^2\right)=0,\\
\nonumber
 &(i\partial_\xi-i\delta\partial_{\tau}-D_2\partial_{\tau\tau})
     U_2+\sqrt{\Delta \beta}\nshgb U_1^2e^{-i\Delta\beta \xi} 
 \\
 &\phantom{i\partial_\xi-D_1}
+2\bar n^2\nkerrb^2 U_2\left(|U_2|^2+\eta\bar n^{-1}|U_1|^2\right)=0,
  \label{eq:shg-bulk-sh-no-kerr}
\end{align}
%\end{subequations}
with $\delta=d_{12}\tin/|\kpp_1|$, %$D_{j}=\sgn(k_{j}^{(2)})/2$,
$D_j=k_j^{(2)}/2|k_{1}^{(2)}|$, 
and $\Delta \beta= \Delta k\ld$. The scaling conveniently gives the SHG
soliton number \cite{moses:2006,bache:2007},
$\nshgb^2=\ld\Ein^2\omega_1^2\deff^2/(c^2 n_{1}n_{2}\Delta k) $, and
the Kerr soliton number $\nkerrb^2=\ld n_{\rm
  Kerr,1}\Ein^2\omega_1/c$.

In the cascading limit $\Delta \beta \gg 1$ the nonlocal approach
takes $U_2(\xi,\tau)= \phi_2(\tau)\exp(-i\Delta\beta\xi)$,
keeping its time dependence
but neglecting the dependence on $\xi$ of $\phi_2$. To do this the
coherence length $\lcoh=\pi/|\Delta k|$ must be the shortest
characteristic length scale in the system, which is true in all
cascaded compression experiments. Discarding the Kerr terms in 
Eq.~(\ref{eq:shg-bulk-sh-no-kerr}) because
$\nkerrb^2 U_2\ll \sqrt{\Delta \beta}\nshgb$, we get an ordinary
differential equation $ D_2 \phi_2''+i\delta \phi_2' - \Delta\beta
\phi_2 =\sqrt{\Delta \beta}\nshgb U_1^2$, with the formal solution
\begin{align}\label{eq:e2-formal}
 \phi_2(\tau) = -\frac{\nshgb}{\sqrt{\Delta\beta}}
 \int_{-\infty}^\infty {\rm d}\tau' R_\pm(\tau') U_1^2(\tau-\tau'). 
\end{align}
According to the sign of the
parameter $s_1=\sgn\left( \Delta\beta/D_2 -
      \delta^2/4D_2^2\right)$, $R_+$ or $R_-$ must be used
%\begin{subequations}
\begin{align}\label{eq:Rplus-dimless}
R_+(\tau) &= \frac{{\tau_2}^2+{\tau_1}^2}{2\tau_1{\tau_2}^2} {\rm
  e}^{-is_2\tau/\tau_2}  \,\,{\rm
e}^{-|\tau|/\tau_1},\\
\label{eq:Rminus-dimless}
R_-(\tau) &= \frac{{\tau_2}^2-{\tau_1}^2}{2\tau_1{\tau_2}^2} 
{\rm e}^{-is_2\tau/\tau_2} \sin(|\tau|/\tau_1),
\end{align}
%\end{subequations}
where $s_2={\rm sgn}(D_2/\delta)$ and $\int_{-\infty}^\infty
{\rm d}\tau R_+(\tau)=1$. Both response functions have an
asymmetric imaginary part due to GVM, causing %and this asymmetry causes 
a Raman effect.  Moreover, $R_+$ is localized in amplitude
(corresponding to the stationary regime), while $R_-$ is purely
oscillating in amplitude (corresponding to the nonstationary regime).
The nonlocal dimensionless temporal scales are
\begin{align}\label{eq:tau-dimless}
%   \tau'_1 = 2|D_2|\left| 4\Delta\beta D_2 -
%       \delta^2\right|^{-1/2}, \quad
  \tau_1 = \left| \Delta\beta/D_2-%\frac{\Delta\beta}{D_2} -
      \tau_2^{-2}\right|^{-1/2}, \quad
\tau_2=2|D_2/\delta|.
%\tau_2=2D_2/\delta.
%   \tau'_1 = \frac{2|D_2|}{\sqrt{ 
%     \left| 4\Delta\beta D_2 -
%       \delta^2\right| }}, \quad\quad
% \tau'_2=\frac{2D_2}{\delta}.
\end{align} 
On dimensional form $t_1 = \tau_1\tin=|2\Delta k/\kpp_2
-t_2^{-2}|^{-1/2}$, $t_2=\tau_2\tin=|\kpp_2/d_{12}|$, and
$\mathcal{R}_\pm=R_\pm/\tin$, which all are independent of
$\tin$.
%\begin{subequations}
% \begin{gather}
% % \label{eq:Rplus-dim}
% % R_+(\tau) = \frac{\Delta k\tau_1}{\kpp_2}\,{\rm
% %   e}^{-i\tau/\tau_2} \,{\rm
% % e}^{-|\tau|/\tau_1}, \\
% % \label{eq:Rminus-dim}
% % R_-(\tau) = \frac{\Delta k \tau_1 }
% %  {\kpp_2}  \,{\rm e}^{-i\tau/\tau_2}
% %  \sin(-|\tau|/\tau_1),
% % \\
% \label{eq:tau-dim}
% %  \tau_1 &= \frac{|\kpp_2|}{\sqrt{|2\Delta k\kpp_2 -d_{12}^2}|},
%   \tau_1 = |\kpp_2||2\Delta k\kpp_2 -d_{12}^2|^{-1/2}, \quad
% \tau_2=\kpp_2/d_{12}.
% \end{gather} 
% %\end{subequations}
The transition between the stationary and nonstationary regimes occurs
when $s_1$ changes sign, which in dimensional units implies that
%\begin{subequations}
\begin{align}\label{eq:stationary}
d_{12}^2&< 2\Delta k k_2^{(2)},\quad {\rm 
  stationary~regime},~R_+, \\\label{eq:nonstationary}
d_{12}^2&>2\Delta k k_2^{(2)},\quad {\rm 
  nonstationary~regime},~R_- .
\end{align}
%\end{subequations}
The transition is independent of $T_0$ but depends on the
GVM, the SH GVD, and the phase mismatch. This central result is
important wherever cascaded quadratic
nonlinear phase shifts are used. It should be
compared with the qualitative arguments of %Liu \textit{et al.}
\cite{liu:1999}, where the stationary regime was $\Delta k<4\pi
  |d_{12}|/ \tin$. 

\begin{figure}[ht]
  \begin{center}
    \centerline{\includegraphics[width=8.3cm]{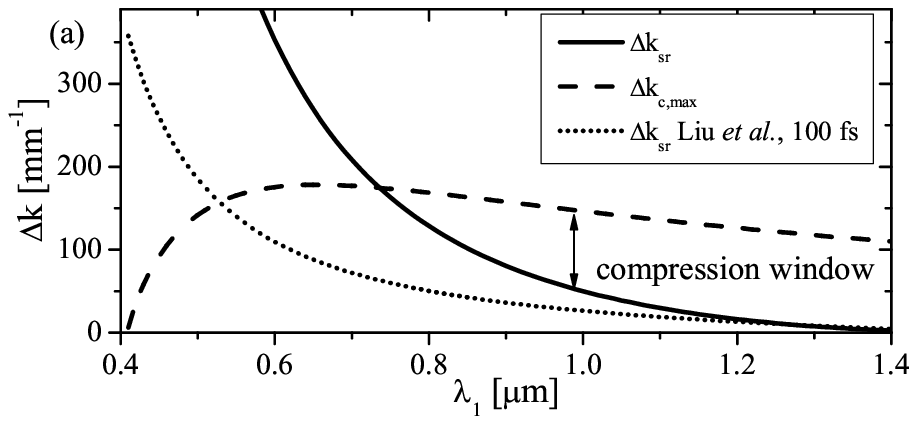}}
    \centerline{\includegraphics[width=8.3cm]{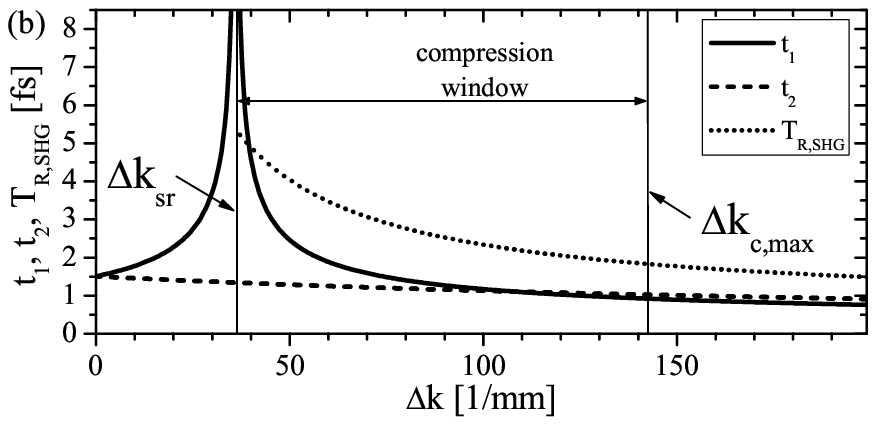}}
\caption{(a) Compression window Eq. (\ref{eq:deltakc}) versus
  $\lambda_1$ in a BBO. Also $\Delta k_{\rm sr}=4\pi
  |d_{12}|/ \tin$ of \cite{liu:1999} for a 100 fs pulse is shown. (b)
  $t_{1,2}$ and $T_{R,\rm SHG}$ versus $\Delta k$ for 
  $\lambda_1=1064$~nm.}\label{fig:bbo}
  \end{center}
\end{figure}

Now, inserting Eq.~(\ref{eq:e2-formal}) into
(\ref{eq:shg-bulk-fh-no-kerr}) gives a dimensionless generalized
nonlinear Schr{\"o}dinger equation 
\begin{multline}
  \label{eq:fh-shg-nlse-nonlocal}
%   i\frac{\partial U_1}{\partial z'}-\frac{1}{2}
%   \frac{\partial^2 U_1}{\partial \tau'^2}+\nkerrb^2U_1|U_1|^2
  [i\partial_{\xi}-D_1\partial_{\tau\tau}]U_1
  +\nkerrb^2U_1|U_1|^2 
\\ 
-\nshgb^2U_1^*\int_{-\infty}^{\infty} {\rm d}\tau'
  R_\pm(\tau')U_1^2(\tau-\tau') =0, 
\end{multline}
In the weakly nonlocal limit of the stationary regime $\tau_{1,2}\ll 1$,
$U_1^2$ is slow compared with $R_+$, so
\begin{multline}
  [i\partial_{\xi}-D_1\partial_{\tau\tau}]U_1
  -(\nshgb^2-\nkerrb^2)U_1|U_1|^2
\\
=is_2\nshgb^2\tau_{R,\rm SHG}|U_1|^2  \partial_\tau U_1,
  \label{eq:fh-shg-nlse-moses}
\end{multline}
where $\int_{-\infty}^\infty {\rm d}t R_+(t)=1$ was used.
%Thus, a similar analysis cannot
%be done in the nonstationary regime because $R_-$ is unbound.  
The dimensionless characteristic time of the nonlocal Raman response is
\begin{align}\label{eq:trSHG}
\tau_{R,\rm SHG}\equiv
\frac{T_{R,\rm SHG}}{\tin}=\frac{4{\tau_1}^2{\tau_2}}
{{\tau_1}^2+{\tau_2}^2}=\frac{2|d_{12}|}{\Delta k\tin}  
\end{align}
Exactly this result has been derived before by perturbative methods
\cite{ilday:2004,bache:2007,moses:2006}; that method therefore amounts
to being in the weakly nonlocal limit of the stationary regime.
Clearly, both Eq.~(\ref{eq:fh-shg-nlse-nonlocal})
and~(\ref{eq:fh-shg-nlse-moses}) have terms reminiscent of the
nonlinear Schr{\"o}dinger equation with a purely cubic nonlinearity
\cite{agrawal:1989}; the GVM-induced nonlocality is similar to the
Raman terms from a delayed cubic response.
%The only difference is that
%Eqs.~(\ref{eq:fh-shg-nlse-nonlocal})-(\ref{eq:fh-shg-nlse-moses})
%include the material Kerr response.

From Eqs.~(\ref{eq:stationary}) and (\ref{eq:fh-shg-nlse-nonlocal}) clean
soliton compression requires, first, being in the stationary regime,
i.e., $\Delta k>\Delta k_{\rm sr}=d_{12}^2/2k_2^{(2)}$.
Second, soliton compression requires
$\neffs=\sqrt{\nshgb^2-\nkerrb^2}>1$ \cite{bache:2007}.  This can be
expressed as $\Delta k<\Delta k_c\equiv \Delta k_{c,\rm
  max}/(1+\nkerrb^{-2})$, where $\Delta k_{c,\rm max}\equiv
\omega_1\deff^2/n_{{\rm Kerr},1}cn_{1}n_{2}$. 
% is the value of $\Delta k_c$ when $\nkerrb\gg 1$. 
To remove the dependence on the FW input
pulse intensity and duration it is convenient to require $\Delta k<
\Delta k_{c,\rm max}$, which is necessary to have
$\neffs>1$. Thus, we obtain a \textit{compression window}:
%\cite{bache:2007}
\begin{align}
\label{eq:deltakc}
&\Delta k_{\rm sr}<\Delta
  k<\Delta k_{c,\rm max}.
%\frac{\Aeff 1}{\Aovl}
%\Delta k_{\rm
%  sr}=\frac{d_{12}^2}{2k_2^{(2)}}, 
%\quad \Delta
%  k_c=\frac{\omega_1}{cn_{1}n_{2}} 
%  \frac{\deff^2}{n_{{\rm  Kerr},1}}\frac{1}{(1+\nkerrb^{-2})}
\end{align}

In Fig.~\ref{fig:bbo}(a) we show the compression window for a
$\beta$-barium-borate (BBO) crystal (see \cite{bache:2007} for BBO
material parameter details).  Notice that the window closes for
$\lambda_1<750$~nm. Opening the compression window here requires a
material with a stronger quadratic nonlinearity, or alternatively a
strong dispersion control, as offered by photonic crystal fibers
\cite{bache:2005a}. At $\lambda_1=800$~nm the window is narrow. In
fact, the compression experiments done at 800 nm
\cite{liu:1999,ashihara:2002} were both in the nonstationary regime,
and were both unable to observe compression to few-cycle pulses. Choosing
$\lambda_1=1064$~nm, Fig.~\ref{fig:bbo}(b) shows the nonlocal time
scales versus $\Delta k$.  At the transition to the nonstationary regime,
$t_1$ diverges while $t_2$ remains low. % Note here that
% Eq.~(\ref{eq:fh-shg-nlse-moses}) reveals a third condition for clean
% compression, namely that the right-hand side remains negligible
% \cite{moses:2006}, implying that $\nshgb^2 \tau_{R,\rm
%   SHG}$ must be small.
%If this is not the case, the solitons will become asymmetric and pulse
%splitting will occur.
\begin{figure}[t]
  \begin{center}
%     \centerline{\includegraphics[width=4cm]{stationary.eps}
%       \includegraphics[width=4cm]{nonstationary.eps}}
    \centerline{\includegraphics[width=8.3cm]{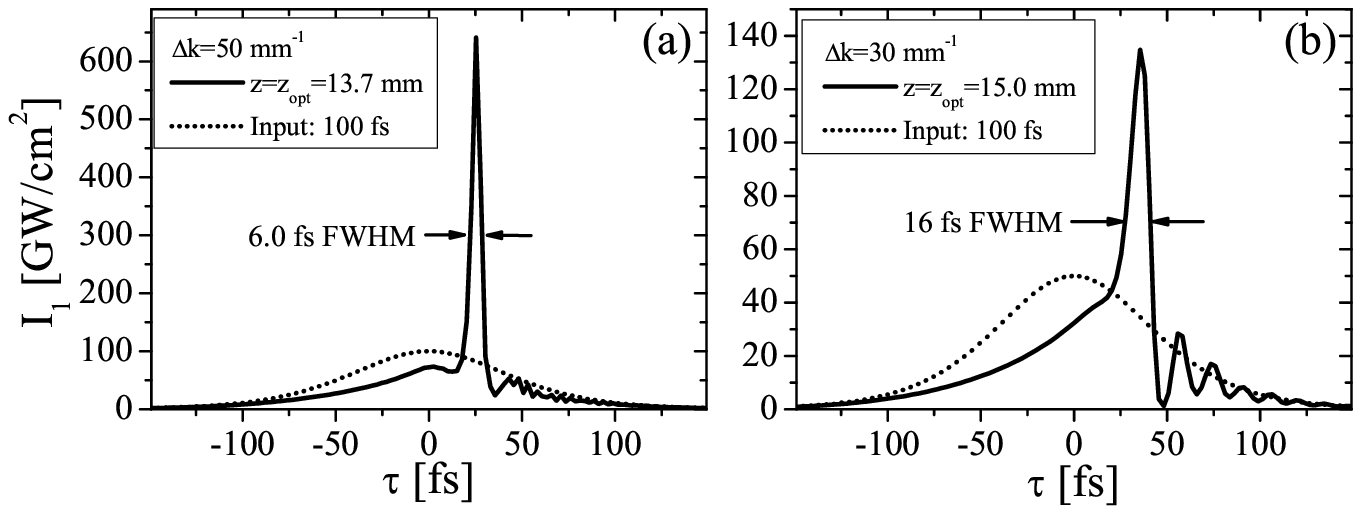}}
%    \centerline{\includegraphics[width=4cm]{dk50Rplus.eps}
%       \includegraphics[width=4cm]{dk30Rminus.eps}}
     \centerline{\includegraphics[width=8.3cm]{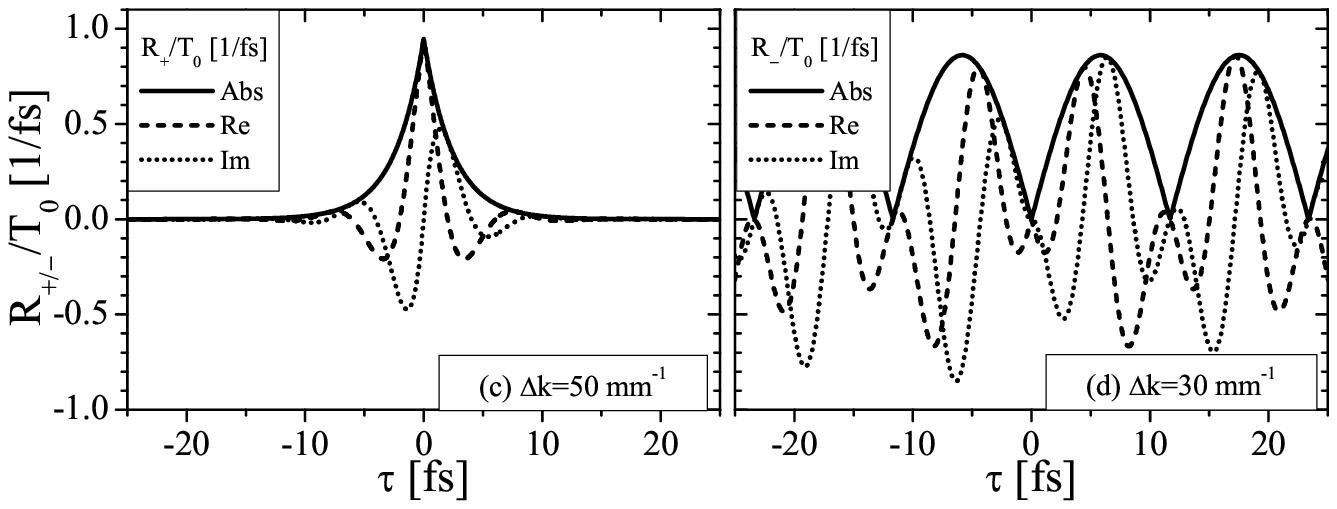}}
     \centerline{\includegraphics[width=8.3cm]{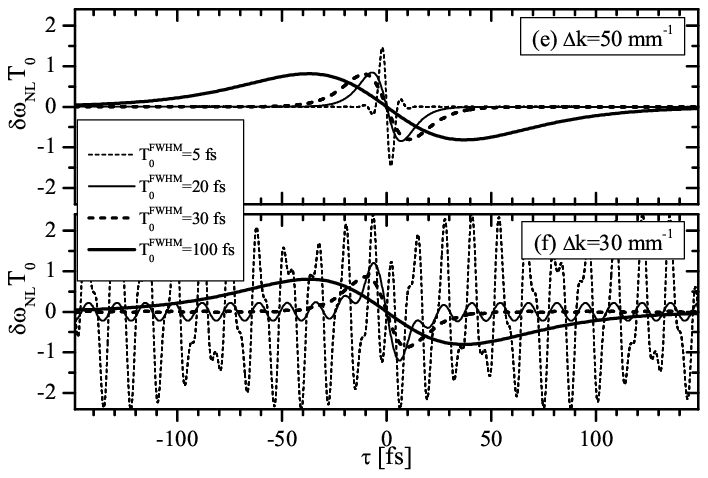}}
\caption{Numerical simulations of soliton compression of a 100 fs
  sech pulse in BBO for $\lambda_1=1064$~nm and $\neffs=5.3$.
  The normalized FW intensity at $z_{\rm opt}$ 
%(distance of optimal
%   compression) 
is shown for (a) $\Delta k
  =50~{\rm mm^{-1}}$ (stationary regime) and (b) $\Delta k
  =30~{\rm mm^{-1}}$ (nonstationary regime).  The response
  functions $R_\pm/\tin$ and $\delta \omega_{\rm NL}\tin$ 
  induced by cascaded SPM for (a) and (b) are shown in (c) and (e) and
  (d) and (f), respectively.}\label{fig:sim}
  \end{center}
\end{figure}

Let us illustrate the two regimes.  Figures~\ref{fig:sim}(a) and
\ref{fig:sim}(b) show two numerical simulations of soliton compression
of a 100 fs pulse at $\lambda_1=1064$~nm in a BBO (the full coupled
equations of \cite{bache:2007} are solved). In Fig.~\ref{fig:sim}(a)
$\Delta k=50~{\rm mm^{-1}}$; our nonlocal theory predicts $\Delta k_{\rm
  sr}=36.0~{\rm mm}^{-1}$, so this is in the
stationary regime. %  The qualitative definition of Liu \cite{liu:1999}
% predicts that it should be deep into the stationary regime. 
Indeed, a symmetric compressed 6 fs pulse is observed. Instead,
changing to the nonstationary regime, $\Delta k=30~{\rm mm^{-1}}$ in
Fig.~\ref{fig:sim}(b), the pulse at $z=z_{\rm opt}$ (optimal
compression point) is very asymmetric and strong pulse splitting
occurs.  Note, the definition in \cite{liu:1999} predicts this
simulation to be in the stationary regime. The nonlocal response
functions are shown in Figs.~\ref{fig:sim}(c) and \ref{fig:sim}(d);
while the nonlocal time scales are clearly quite similar for both
examples, the different shapes of the response functions imply a very
different impact on the pulse dynamics. This can be understood by
using Eq.~(\ref{eq:fh-shg-nlse-nonlocal}) to calculate the chirp
$\delta \omega_{\rm NL}$ induced by SPM from the cascaded SHG process
of a sech input pulse (calculations as in \cite{agrawal:1989}, Chap.
4).  Figures~\ref{fig:sim}(e) and \ref{fig:sim}(f) show $\delta
\omega_{\rm NL}$ for $z=L_{\rm SHG}$ (where $\nshgb^2\equiv\ld/L_{\rm
  SHG}$ \cite{bache:2007}).  In the stationary case,
Fig.~\ref{fig:sim}(e), SPM induces a linear negative chirp on the
central part of the pulse because $R_+$ is localized, except for very
short pulses, $\tin\sim t_1$, where $R_+$ becomes nonlocal. In the
nonstationary case, Fig.~\ref{fig:sim}(f), SPM induces a linear chirp
only for long pulses $>30$ fs, while shorter pulses also have strong
chirp induced in the wings because of the oscillatory character of
$R_-$; this explains the trailing pulse train in
Fig.~\ref{fig:sim}(b). Thus, the nonstationary response can be
nonlocal even for $\tin\gg t_{1,2}$. Note, in Fig.~\ref{fig:sim}(b)
very short pulses ($T_0\sim t_{1,2}$) are \textit{positively} chirped
in the central part (equivalent to a chirp induced by a self-focusing
nonlinearity), making few-cycle compressed pulses impossible in the
nonstationary case.

% To conclude GVM induces asymmetric nonlocal responses that 
% accurately explain the stationary and nonstationary regimes in
% cascaded quadratic soliton compressors. The transition is determined
% by the GVM, the phase mismatch and the SH GVD. While the nonlocal
% response functions are independent on the input pulse length, the
% input pulse duration relative to the nonlocal time scales does play an
% important role. These results offer new insight into the physics
% of this soliton compressor.
To conclude we showed that GVM induces asymmetric nonlocal Raman
responses that accurately explain the stationary and nonstationary
regimes in cascaded quadratic soliton compressors. The nature
of the response functions and their degree of nonlocality relative to
the input pulse length is vital for the resulting compression. The
theory offers new insight into the physics of this soliton compressor.

M. Bache was supported by The Danish Natural Science Research
Council Grant 21-04-0506.

%\bibliography{d:/Projects/Bibtex/literature}
%\bibliographystyle{D:/Projects/localtexmf/OSA/ol}
%\bibliographystyle{D:/Projects/localtexmf/prsty}
%\bibliographystyle{apsrev}

%\pagebreak

\end{document}